\documentclass[twocolumn]{27onr_art}
\usepackage{ulem}
\usepackage{harvard}
\usepackage{graphicx}

\pdfoutput = 1

\def\be{\begin{eqnarray}}
\def\ee{\end{eqnarray}}
\def\benl{\begin{eqnarray*}}
\def\eenl{\end{eqnarray*}}

\newcommand{\nwc}{\newcommand}
\nwc{\bm}{\boldmath}
\nwc{\m}{\mbox}
\nwc{\ubm}{\unboldmath}
\nwc{\bmU}{\m{\bm$U$\ubm}}
\nwc{\bmX}{\m{\bm$X$\ubm}}
\nwc{\bmu}{\m{\bm$u$\ubm}}
\nwc{\bmx}{\m{\bm$x$\ubm}}
\nwc{\bmz}{\m{\bm$z$\ubm}}
\nwc{\bmv}{\m{\bm$v$\ubm}}
\nwc{\bmw}{\m{\bm$w$\ubm}}
\nwc{\bmW}{\m{\bm$W$\ubm}}
\nwc{\bmn}{\m{\bm$n$\ubm}}
\nwc{\bmG}{\m{\bm$G$\ubm}}
\nwc{\bmF}{\m{\bm$F$\ubm}}
\nwc{\bmI}{\m{\bm$I$\ubm}}
\nwc{\bmN}{\m{\bm$N$\ubm}}
\nwc{\bmP}{\m{\bm$P$\ubm}}
\nwc{\bmcalP}{\m{\bm $\cal P$\ubm}}
\nwc{\bmV}{\m{\bm$V$\ubm}}
\nwc{\bmS}{\m{\bm$S$\ubm}}

\begin{document}

\title{An Integrated Experimental and Computational Investigation into the Dynamic Loads and Free-surface 
    Wave-Field Perturbations Induced by Head-Sea Regular Waves on a 1/8.25 Scale-Model of the R/V ATHENA}

\author{Toby Ratcliffe$^1$, Lisa Minnick$^1$, Thomas T. O'Shea$^2$, Thomas Fu$^1$, Lauren Russell$^1$,  and Douglas G. Dommermuth$^2$}

\affiliation{\small $^1$Naval Surface Warface Center,
\\ 9500 MacArthur Blvd., West Bethesda, MD 20817
\\ $^2$Naval Hydrodynamics Division, Science Applications International Corporation,
\\ 10260 Campus Point Drive, MS 35, San Diego, CA  92121}

\maketitle

\begin{abstract}
A 1/8.25 scale-model of the U.S. Navy Research Vessel ATHENA was tested in regular head-sea waves to obtain data for validation of computational fluid dynamics (CFD) predictive tools.  The experiments were performed in the David Taylor Model Basin at the Naval Surface Warfare Center (NSWC). With the model towed fixed in head-seas, horizontal and vertical loads on the model were obtained at two Froude numbers, $F_r=0.25$ and $F_r=0.43$. The model was run at two conditions of head-sea wavelengths corresponding to  $\lambda=2L_o$ and  $\lambda=1/2L_o$ with $H/\lambda=0.03$, where $L_o$ is the length of the model and $H=2 a$ is the wave height. The wave field perturbations induced by the head-sea waves were quantified from free-surface images generated by a laser light sheet. Predictions of the horizontal and vertical loads on the model in regular head sea waves were made with the Numerical Flow Analysis (NFA) code.  Numerical predictions of the wave-field perturbations were compared with the experimental data and the correlation coefficients have been computed.  
\end{abstract}

\section{Introduction}

\begin{figure}
\begin{center}
\includegraphics[width=0.9\linewidth]{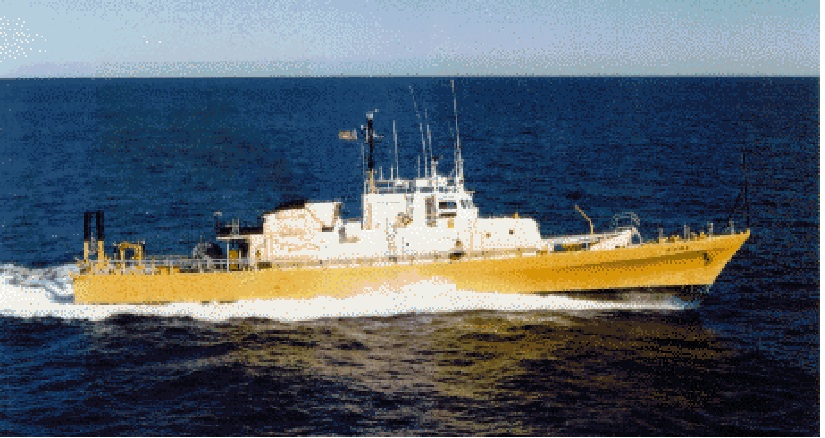} \\
\caption{\label{athena} R/V ATHENA.}
\end{center}
\end{figure}

NSWC Model 5365 is a 1/8.25 scale model of the \mbox{U. S.} Navy Research Vessel ATHENA. The R/V ATHENA is a converted PG-84 Asheville-class patrol gunboat which is operated out of Naval Surface Warfare Center Panama City Division as a high-speed research vessel. A photograph of the R/V ATHENA  is shown in Figure \ref{athena}. In conjunction with the model-scale experiments presented in this paper, the Office of Naval Research (ONR) funded a set of full-scale experiments aimed at quantifying the breaking bow-wave of this vessel \cite{fu04}. 

Model  5365 has recently been used  by ONR as a candidate hull form to evaluate predictive computational fluid dynamics (CFD) codes \cite{wilson06}. Two features of this hull design have contributed to increasing interest in its use for CFD validation:  the large speed range of the model, corresponding to 6 - 35 knots full scale, representing a Froude number range from 0.14 - 0.83, including displacement and planing speeds; and the transom-stern geometry of the hull form, which is realistic of a naval combatant.

A large quantity of data has been obtained for the model in calm water including resistance, sinkage and trim, and longitudinal wave cuts \cite{ratcliffe07b}. The natural extension of this data set would be to obtain a comprehensive set of loads data for a fixed sinkage and trim condition over a range of incoming wave heights. Having access to this data set would allow CFD code developers to validate the model loads with their code before extending the predictions to the free-model motion in waves. 

\section{Description of laboratory experiments}

\subsection{Model characteristics}
	
Model 5365 was built in 1979 at the Naval Surface Warfare Center, and was fabricated out of wood. The linear scale ratio is 8.25.   Model and full-scale ship characteristics are shown in Table \ref{tab:scale}.  

\begin{table}
\begin{tabular}{|l|c|c|}\hline
	 & Model Scale	& Full Scale \\ \hline Displacement 	& 397 kg 	& 229 metric tons \\
	 & (875 lbs) & (225 long tons) \\ \hline Draft (hull)	& 0.19 m (0.618 ft) &	1.6 m (5.1 ft) \\ \hline Maximum Beam &	0.84 m (2.74 ft) & 	6.9 m (22.6 ft) \\ \hline Transom Beam	 & 0.70 m (2.3 ft) & 	5.8 m (19.0 ft) \\ \hline LBP 	& 5.69 m (18.67 ft) &	46.9 m (154.0 ft) \\ \hline\end{tabular}
\caption{Model 5365 and Full-Scale (R/V ATHENA) Hull-Form Characteristics \label{tab:scale}}
\end{table}

\subsection{The towing basin and wave-generation capability}

The model was tested in the Carriage 2 Basin at the Naval Surface Warfare Center, Carderock, Maryland, USA. This deep water basin is 6.7 meters deep, approximately 575 meters long and 15.5 meters wide. A pneumatic wavemaker is located at one end, and a wave-absorbing beach at the other.  Behind a moveable section of the beach is a fitting-out dry dock.   The pneumatic wavemaker is a 15.5 meter wavemaker dome divided into two equal length sections, connected to a centrifugal type blower. The blower is driven by a 112 kW variable speed DC electric motor.  The wavemaker can generate regular waves from 1.5 to 12.2 meters in length, with corresponding peak wave heights of 10 to 61 cm.  

\subsection{Instrumentation}

	Instrumentation used included Kistler 6-component force and moment gages to measure the forces on the model, sonic wave-height probes to measure the frequency and amplitude of the incoming waves, and a quantitative visualization (Q-Viz) laser light sheet system to measure the near-field free-surface wave field around the model. The data were collected using LabView software, a National Instruments product.  In conjunction with this collection software, National Instruments hardware, consisting of a CPU and  A/DÕs was also used.  Details of the specific instrumentation are included in the following sub-sections.\subsubsection{Kistler 6-component force gauge}

	The Kistler force gages are piezoelectric force sensors with integrated charge amplifier electronics,  sandwiched inside stainless steel plates and pre-loaded according to the manufacturerÕs specifications. The gages were located in the model at two positions. The forward Kistler gage was located at  $x/ L_o$ =0.25 and the aft gage at $x/ L_o$ =0.81.  The primary data obtained from the gages for this experiment were average and peak  x-force and z-force.  This data was collected at 100 Hz. \subsubsection{Sonic wave height probes}

 	Wave height measurements were taken using Senix TS-15S-IV ToughSonic distance sensors.  These sensors emit an ultrasonic pulse that bounces off of the water and the return pulse is then read using a piezo-electric element.  Using the speed of sound, the sensor is able to calculate the distance to the water surface. The ToughSonics can measure distances from 10 to 360 inches, with an accuracy of 0.12 cm.   The sampling rate of the ToughSonics used in these experiments was set to 10 Hz. \subsubsection{Quantitative free-surface visualization}

A non-intrusive optical technique, Quantitative Visualization (QViz), has been developed to measure the free-surface disturbances occurring in regions commonly inaccessible to more traditional measurement methods, i.e. near wake flows, bow sheets and breaking waves.  These regions are generally difficult to quantify due to the multiphase aspect of the flow as well as their unsteady nature.  However, the unsteady surfaces, droplets, and bubbles in these regions are effective scatterers and allow for optical imaging of the deformations of the surface. This technique has been used extensively to measure free-surface elevations and breaking waves \cite{fu03,karion03}.
	
\begin{figure}
\begin{center}
\includegraphics[width=0.9\linewidth]{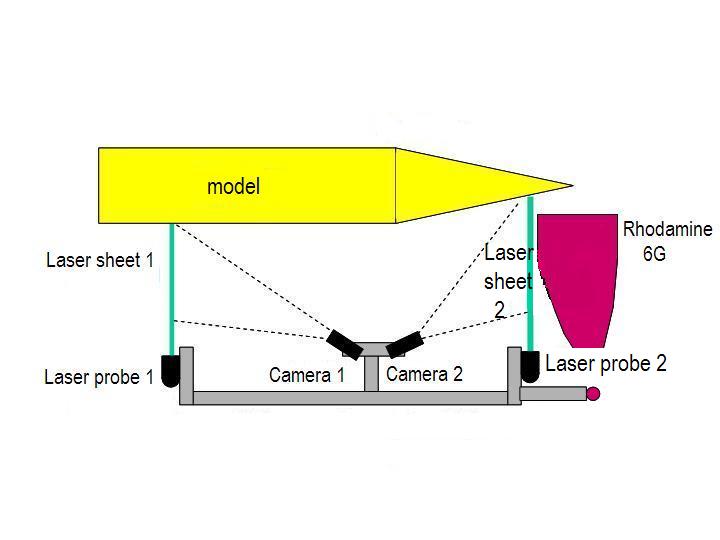} \\
\caption{\label{qviz} Schematic of Q-Viz system.}
\end{center}
\end{figure}

\begin{figure}
\begin{center}
\includegraphics[width=0.9\linewidth]{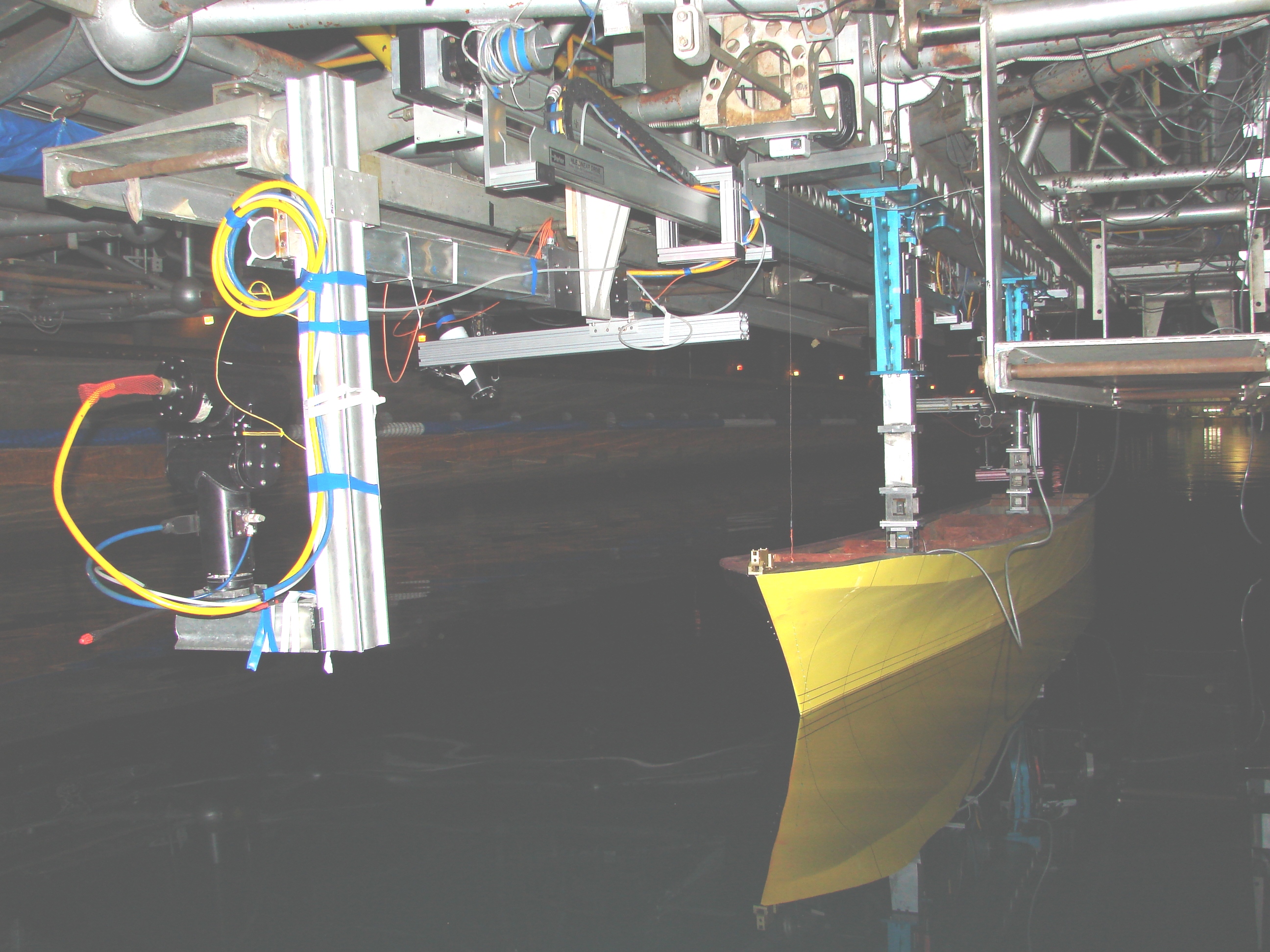} \\
\caption{\label{modelbow} Q-Viz instrumentation as viewed from the bow of the model.}
\end{center}
\end{figure}

\begin{figure}
\begin{center}
\includegraphics[width=0.9\linewidth]{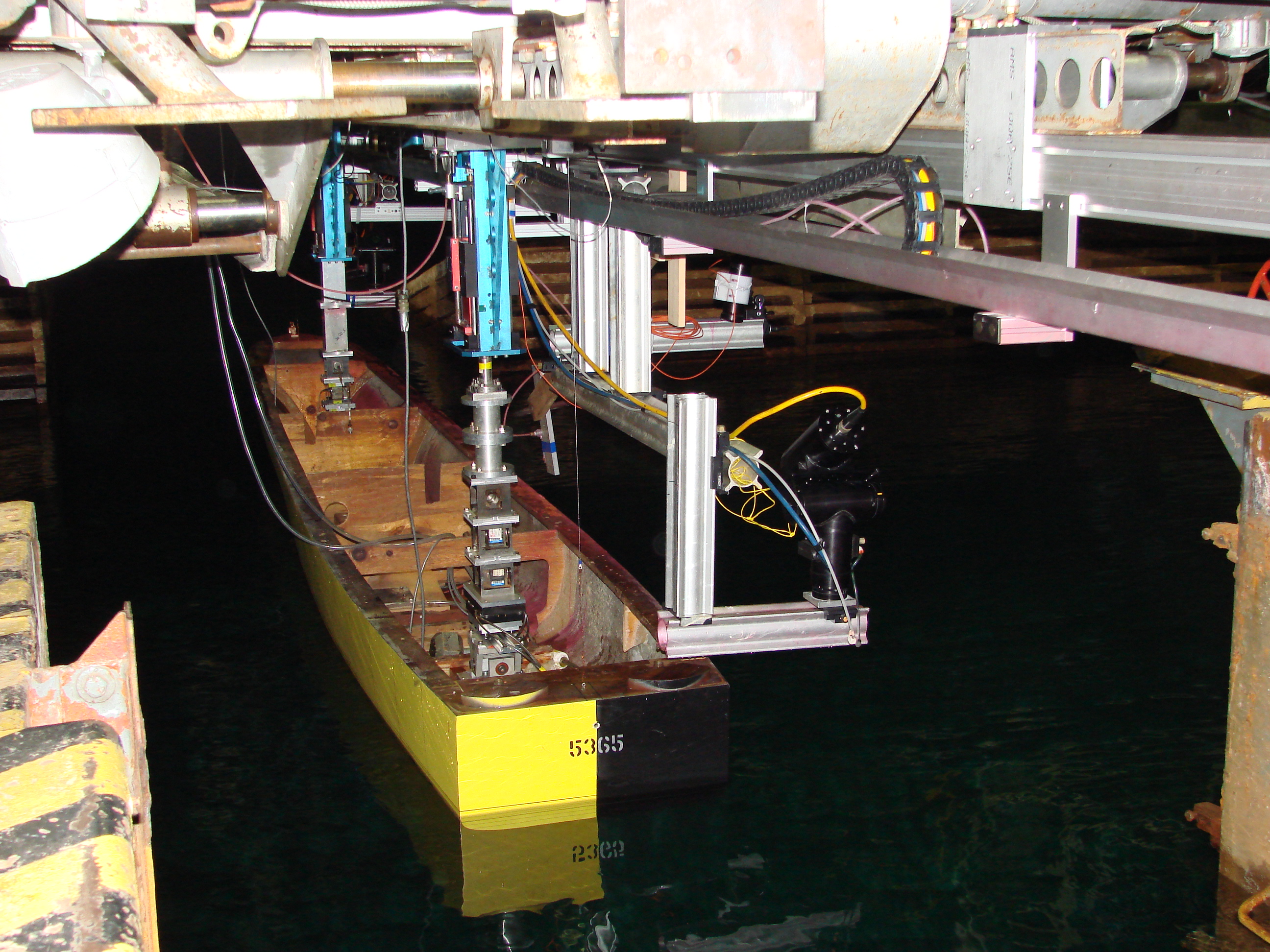} \\
\caption{\label{modelstern} Q-Viz instrumentation as viewed from the stern of the model.}
\end{center}
\end{figure}

The QViz system consists of a continuous wave laser and optics to create a steerable light sheet.  The light sheet and collection optics are mounted at a specific orientation relative to the flow.  The laser beam is coupled into a fiber-optically fed light probe.  For the current set up, two light sheets were generated perpendicular to the model center line and the free-surface at two different axial locations (referred to as the forward location and the aft location).  A digital video camera was directed towards each light sheet.  A schematic of the Q-Viz system is shown in Figure \ref{qviz} and photographs of the model and camera system, as viewed from the bow and stern, are shown in Figures \ref{modelbow} and \ref{modelstern}. 

Images from each camera were collected at 30 frames/second using two National Instruments frame-grabber boards and two personal computers (one for each camera).  An image analysis program was developed at NSWCCD using National Instruments Labview software and an image processing toolbox to extract the surface profile information.  Sequential images (usually 30 images, representing one second of data) were analyzed and then averaged together, providing a time-averaged profile. A more detailed description of the system and itÕs capabilities can be found in \citeasnoun{rice04}.\subsection{Experimental conditions}
Prior to the in-waves experiments, calm water data were obtained on the model at a free-to-sink-and-trim condition over a Froude number range from Fr=0.14-0.84. From this data, the fixed-trim conditions could be specified for the experiments in head-sea waves \cite{ratcliffe07b}. 	
The model was then run at two speed-determined fixed-trim conditions for two wave conditions and at two Froude numbers. These conditions are summarized in Table \ref{tab:matrix}.  The wave conditions were chosen such that the wave slope value, $H/\lambda$, was equal to 0.03 for all the waves. 

\begin{table}
\begin{tabular}{|p{0.2\linewidth}|p{0.2\linewidth}|p{0.2\linewidth}|p{0.2\linewidth}|}\hline
Froude Number &	Head-Sea Wave Length	& Target Wave Height (peak-to-trough) & Target Wave Period \\ \hline0.25 & $L_o/2$ & 9.14 cm &	1.3 sec \\ \hline0.25	& $2L_o$ & 32.8 cm	& 2.7 sec \\ \hline0.43 & $L_o/2$ & 9.14 cm	& 1.3 sec \\ \hline0.43	& $2L_o$ & 32.8 cm & 2.7 sec \\ \hline
\end{tabular}
\caption{Test Matrix for Model 5365 in Head Sea Waves \label{tab:matrix}}
\vspace{0.5in}
\end{table}

\subsection{Data analysis}

\subsubsection{Fixed trim data}

	The data analysis software, MATLAB, was used to analyze the fixed-trim data. Maximum amplitudes of the wave height, drag force and vertical force were found for each data spot.  Drag and z-forces, measured at the separate forward and aft gages were added to produce total forces on the model. 
	
\begin{table}
\begin{tabular}{|l|p{0.25\linewidth}|p{0.25\linewidth}|p{0.25\linewidth}|}\hline
Fr &	Wave Amplitude & Average Peak  Vertical Force & Average Peak Drag Force \\
    & (cm) & (newtons) & (newtons) \\ \hline			0.25	& 9.65 & 76.5 & 71.2 \\ \hline0.25	& 30.73 & 2687.6 &	387 \\ \hline0.43	& 9.65 &	50.5 & 	182.4 \\ \hline0.43	& 33.52 &	2995.3 &	547.1 \\ \hline
\end{tabular}
\caption{Summary of Peak-to-Peak Vertical and Horizontal Forces Obtained on Model 5365 in Head-Sea Waves \label{tab:summary}}
\end{table}

The data was analyzed to compute the drag force and vertical force at each model condition. A summary of the experimentally measured forces is presented in Table \ref{tab:summary}. This table shows that the peak vertical force is almost constant with similar wave amplitudes, regardless of model speed. The average peak drag force, however, increases with both increasing speed and increasing wave amplitude. \subsubsection{Quantitative free-surface visualization data}
	
Data from the forward camera/light sheet were collected at 28 longitudinal locations along the model ranging from 5.6 inches forward of the FP to 145.6 inches aft of the FP. The locations were equally spaced at 5.6 inches apart. The raw data for each position consisted of consecutive frames that represent the free surface at a given moment in time. The data are first analyzed to produce a phase-averaged profile. For each frame the average wave height is determined and a time trace of the mean surface is produced. The time trace mean is then used to determine the perceived period of the waves from the model perspective. The dominant frequency of the time trace is found using an FFT function. Using the period determined by the FFT, the time trace is analyzed to determine how many wave encounters were observed. The time series is then split up into sections so that each wave period in the trace is divided into N sections, each comprising of a fraction (1/N) of a period. For the larger amplitude, longer wavelength, $\lambda=2L_o$ waves, each wave period was divided into 80 sections. The frame number(s) corresponding to each of the sections are then determined and the free-surface profiles of the frame numbers for each section are averaged together for each encounter. Once data from each position has been phase-averaged, the results can be used to plot the entire free surface at a given instant in time.
\section{Description of numerical simulations}
\subsection{Introduction to NFA}

Predictions of the horizontal and vertical loads on the model in regular head sea waves were made with the Numerical Flow Analysis (NFA) code at Fr=0.43.  This code provides turnkey capabilities to model breaking waves around a ship, including both plunging and spilling breaking waves, the formation of spray, and the entrainment of air.   Cartesian-grid methods are used to model the ship hull and the free surface.  Following \citeasnoun{puckett97} and \citeasnoun{sussman01}, a cut-cell method is used to enforce free-slip boundary conditions on the hull.  A surface representation of the ship hull is used as input to construct fractional areas and volumes.  The interface capturing of the free surface uses a second-order accurate, volume-of-fluid technique.   At each time step, the position of the free surface is reconstructed using piece-wise planar surfaces as outlined in \citeasnoun{rider94}.  A second-order, variable-coefficient Poisson equation is used to project the velocity onto a solenoidal field thereby ensuring mass conservation. A multigrid method is used to solve the Poisson equation.   Details of a similar projection operator are provided in \citeasnoun{puckett97}. The convective terms in the momentum equations are accounted for using a slope-limited, third-order QUICK scheme as discussed in \citeasnoun{leonard97}. The governing equations are solved using a domain decomposition method.  Communication between processors on the Cray T3E is performed using MPI.   The CPU requirements are linearly proportional to the number of grid points and inversely proportional to the number of processors.

\subsection{Numerical Formulations} \label{sec:numerics}

\citeasnoun{dommermuth07} and \citeasnoun{dommermuth08} provide details of the NFA formulation.  Here, we highlight the formulation of a wavemaker.   Let  $\eta(x,t)$ denote the free-surface elevation as function of position  $x$ and time $t$, then
\begin{eqnarray}
\eta(x,t) & = & a f(t) \cos(k x + \sigma t) \nonumber \\
              & +  & \frac{a}{2} k a f(t) \cos(2 k x + 2 \sigma t)  \;\; ,
\end{eqnarray}
\noindent where $a$  is the wave amplitude,  $k$ is the wavenumber, and  $\sigma$ is the encounter frequency.  The preceding formula is accurate to second order in wave steepness.    $f(t)$ is an adjustment factor that slowly ramps up the wave amplitude \cite{dommermuth08}.  The encounter frequency is a function of the intrinsic wave frequency $\omega$, the wavenumber, and the speed of the free-stream current $U_o$ :
\begin{eqnarray}
\sigma=\omega - k U_o(t)   \;\; ,
\end{eqnarray}
\noindent where the normalized intrinsic wave frequency is
\begin{eqnarray}
\omega^2=\frac{k}{F_r^2} \tanh(k d)  \;\; ,
\end{eqnarray}
\noindent where $d$ is the water depth.  The speed of the ship is slowly ramped up from rest using the adjustment factor.
\begin{eqnarray}
U_o(t)=-f(t)  \;\; .
\end{eqnarray}
For $z \leq \eta$ , the horizontal and vertical components of the water-particle velocity are
\begin{eqnarray}
u(x,z,t) & = & -a \omega f(t) \frac{\cosh(k (z+d))}{\sinh(k d)} \cos(k x +\sigma t)  \nonumber \\
w(x,z,t) & = & -a \omega f(t) \frac{\sinh(k (z+d))}{\sinh(k d)} \sin(k x +\sigma t)  \;\; . \nonumber \\
\end{eqnarray}
For $z>\eta$ , the horizontal and vertical components of the air-particle velocity are
\begin{eqnarray}
u(x,z,t) & = &  a \omega f(t) \frac{\cosh(k (z-h))}{\sinh(k h)} \cos(k x +\sigma t)  \nonumber \\
w(x,z,t) & = &  a \omega f(t) \frac{\sinh(k (z-h))}{\sinh(k h)} \sin(k x +\sigma t)  \;\; , \nonumber \\
\end{eqnarray}
\noindent where $h$ is the height of the air above the free surface.  Using this formulation, the horizontal water-particle velocity is discontinuous across the air-water interface, and the vertical water-particle velocity is continuous across the air-water interface.   The proceeding expressions for the free-surface elevations and the water and air-particle velocities are imposed in a region ahead of the bow.    Aft of this region, the fluid motion is free to evolve according to the governing equations.

\section{Results}

\subsection{Wave forces}

A three-dimensional numerical simulation was performed using 850x192x128= 20,889,600 grid points, 5x8x4=160 sub-domains, and 160 nodes, on a Cray XT3.  The length, width, depth, and height of the computational domain are respectively 4.0, 1.0, 1.0, 0.5 ship lengths ($L_o$).  Grid stretching is employed in all directions.  The smallest grid spacing is 0.0020  near the ship and mean waterline, and the largest grid spacing is 0.020  in the far field.  The Froude number is $F_r=0.43$. Two incident wavelengths are considered: $\lambda=2L_o$ and  $\lambda=1/2L_o$.   In both cases, the wave steepness is $H/\lambda=0.03$, where $H=2 a$ is the wave height. The equations for the wavemaker are imposed ahead of the ship over the range $1\leq x \leq 1.5$. Initial transients are minimized by slowly ramping up the free-stream current and the incident wave amplitude. For this simulation, the non-dimensional time step is t=0.0005.  The numerical simulation runs 10,000 time steps corresponding to 5 ship lengths.  Each simulation requires about 80 hours of wall-clock time.

\begin{figure}[!h]
\begin{center}
\includegraphics[width=0.9\linewidth]{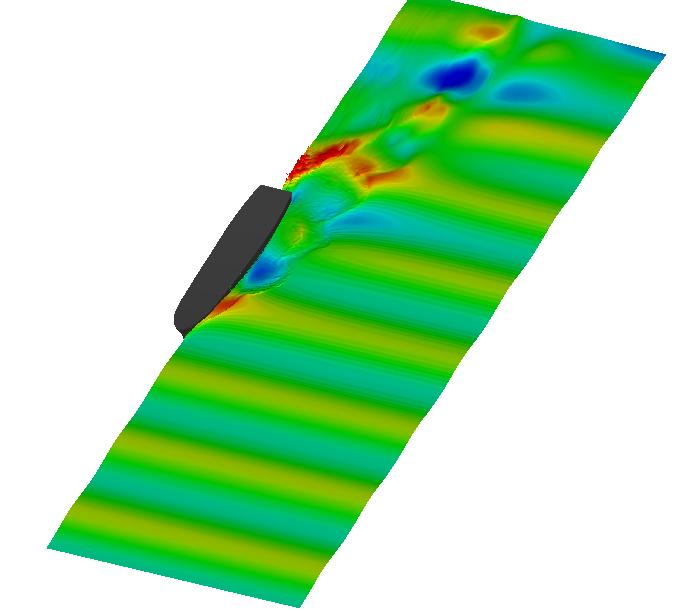}
\caption{\label{perspective_sml} Wave elevation.  $\lambda/L_o=1/2$.}
\end{center}
\end{figure}

\begin{figure}[!h]
\begin{center}
\includegraphics[width=0.9\linewidth]{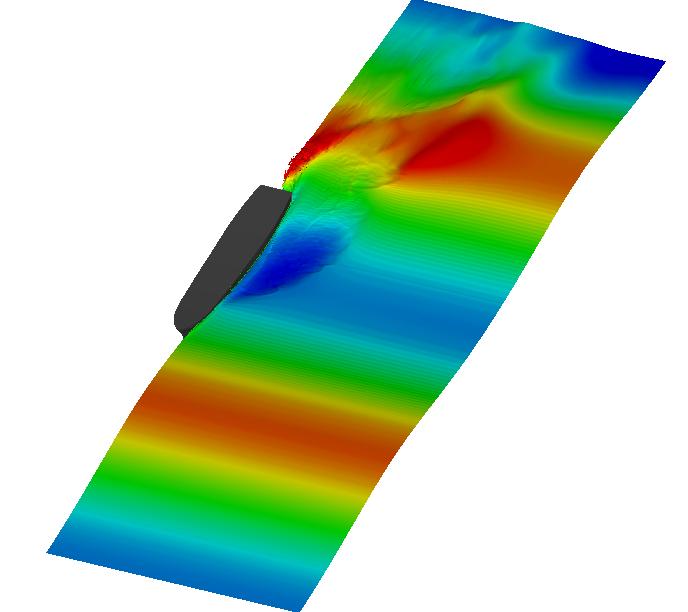}
\caption{\label{perspective_big} Wave elevation.  $\lambda/L_o=2$.}
\end{center}
\end{figure}

Figures \ref{perspective_sml} \& \ref{perspective_big} show perspective views of the predicted free-surface elevations at time t=5  for the two wave amplitudes which were modeled.

%\begin{figure*}
%\begin{center}
%\begin{tabular}{lclc}
%(a)  \vspace{-7pt} & & (b)  \vspace{-7pt} & \\
%& \includegraphics[width=0.4\linewidth]{sml.jpg} & & \includegraphics[width=0.4\linewidth]{big.jpg} \\
%\end{tabular}
%\caption{\label{perspective} Wave elevation.  (a) $\lambda/L_o=1/2$. (b) $\lambda/L_o=2$.}
%\end{center}
%\end{figure*}

\begin{figure}
\begin{center}
\includegraphics[width=0.9\linewidth]{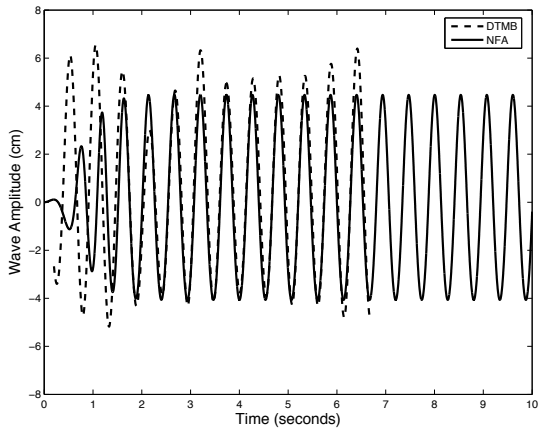}
\caption{\label{elevate_sml} Wave elevation at $x=1.5$.  $\lambda/L_o=1/2$.}
\end{center}
\end{figure}

\begin{figure}
\begin{center}
\includegraphics[width=0.9\linewidth]{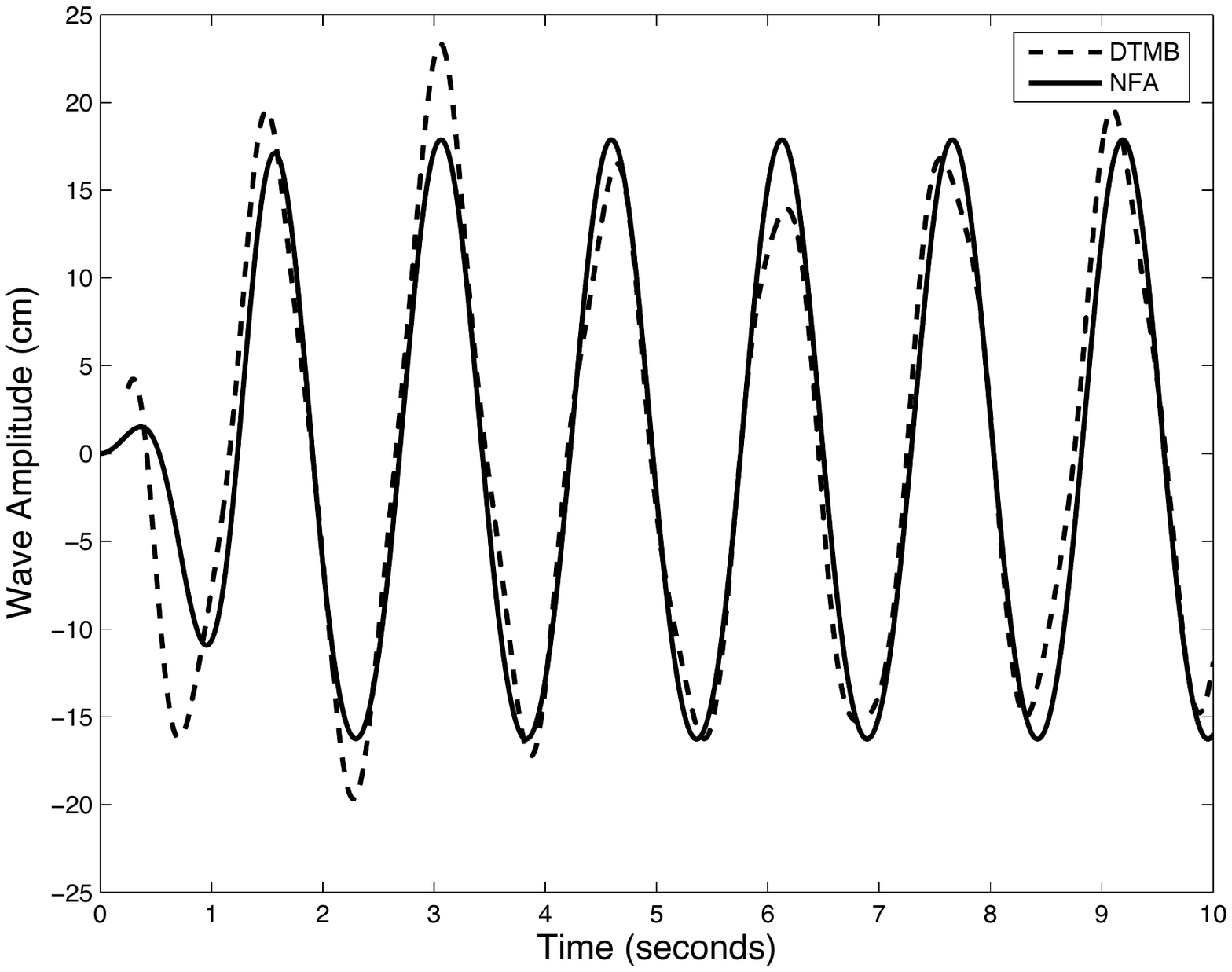}
\caption{\label{elevate_big} Wave elevation at $x=1.5$.  $\lambda/L_o=2$.}
\end{center}
\end{figure}

The predicted and measured free-surface elevations as a function of time are shown in Figures \ref{elevate_sml}  \& \ref{elevate_big} .  The predicted free-surface elevations are ramped up to their full height.  This minimizes transients associated with starting up the numerical wavemaker.  The measured free-surface elevations show slight irregularities that are associated with limitations with the wavemaker at NSWCCD.

%\begin{figure*}
%\begin{center}
%\begin{tabular}{lclc}
%(a)  \vspace{-7pt} & & (b)  \vspace{-7pt} & \\
%& \includegraphics[width=0.4\linewidth]{sml_amp.jpg} & & \includegraphics[width=0.4\linewidth]{big_amp.jpg} \\
%\end{tabular}
%\caption{\label{elevate} Wave elevation at $x=1.5$.  (a) $\lambda/L_o=1/2$. (b) $\lambda/L_o=2$.}
%\end{center}
%\end{figure*}

\begin{figure}
\begin{center}
\includegraphics[width=0.9\linewidth]{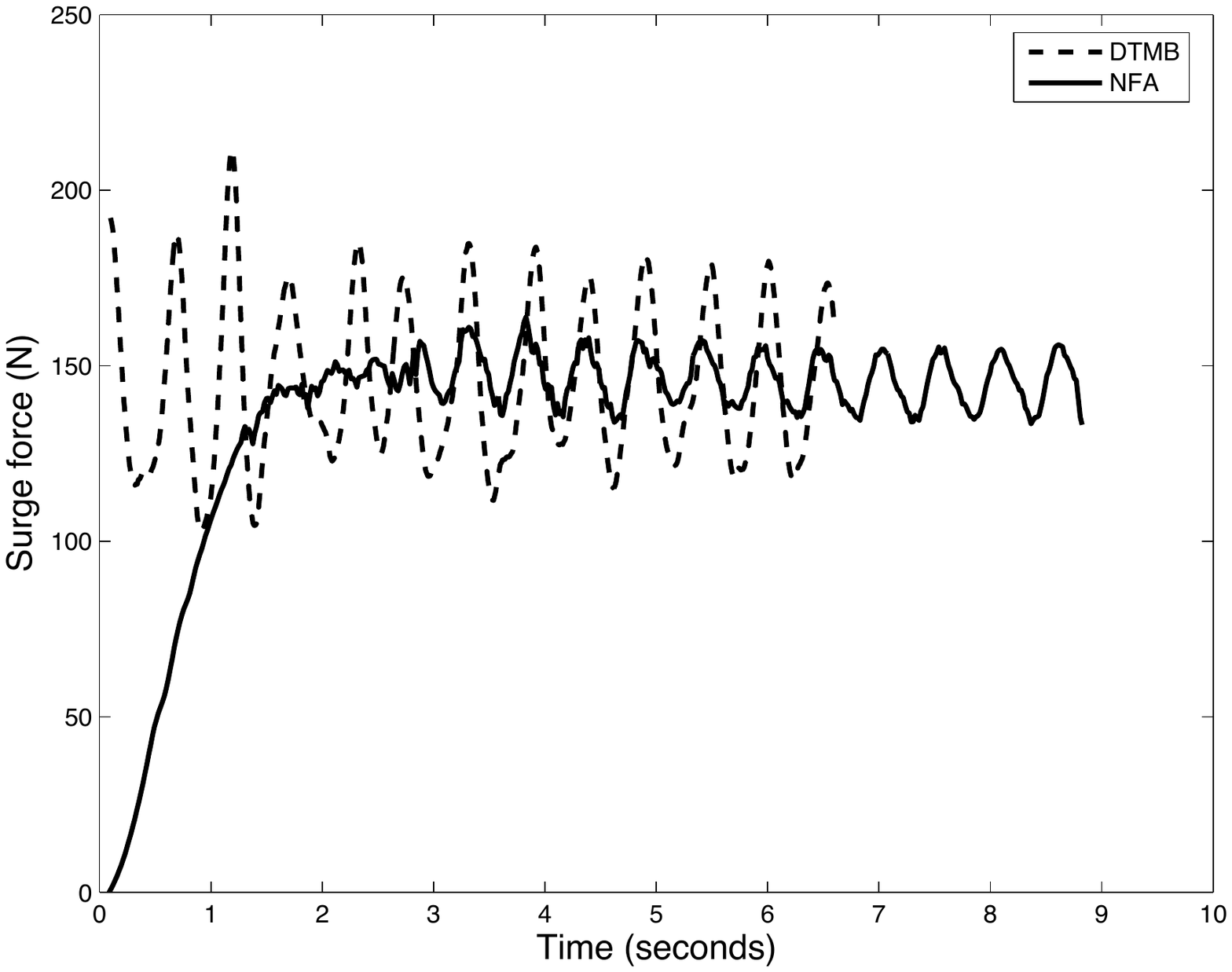}
\caption{\label{drag_sml} Drag (surge) force.  $\lambda/L_o=1/2$.}
\end{center}
\end{figure}

\begin{figure}
\begin{center}
\includegraphics[width=0.9\linewidth]{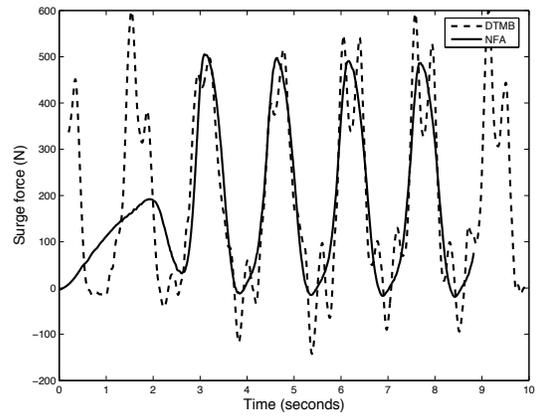}
\caption{\label{drag_big} Drag (surge) force.  $\lambda/L_o=2$.}
\end{center}
\end{figure}

Figures \ref{drag_sml}  \& \ref{drag_big} show the predicted drag force compared to measurements. In the case of the numerical simulations, the drag is initially zero because the model is ramped up to full speed from zero forward speed.   The primary harmonics that are evident in the plots are due to the incident wave forces.   We speculate that the higher harmonics that are evident in the laboratory results, which are especially evident for the longer wave, are due to vibrations in the structure that is used to restrain the model.   In general, numerical predictions and laboratory measurements agree well, especially for the long wave length case. 

%\begin{figure*}
%\begin{center}
%\begin{tabular}{lclc}
%(a)  \vspace{-7pt} & & (b)  \vspace{-7pt} & \\
%& \includegraphics[width=0.4\linewidth]{sml_surge.jpg} & & \includegraphics[width=0.4\linewidth]{big_surge.jpg} \\
%\end{tabular}
%\caption{\label{drag} Drag (surge) force.  (a) $\lambda/L_o=1/2$. (b) $\lambda/L_o=2$.}
%\end{center}
%\end{figure*}

\begin{figure}
\begin{center}
\includegraphics[width=0.9\linewidth]{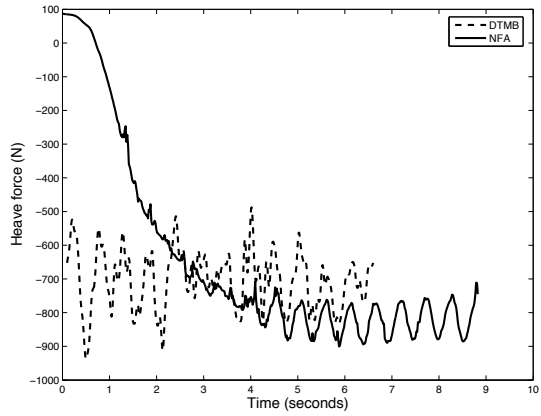}
\caption{\label{heave_sml} Vertical (heave) force.  $\lambda/L_o=1/2$.}
\end{center}
\end{figure}

\begin{figure}
\begin{center}
\includegraphics[width=0.9\linewidth]{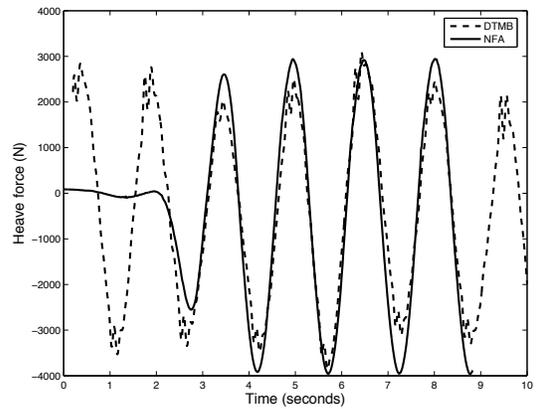}
\caption{\label{heave_big} Vertical (heave) force.  $\lambda/L_o=2$.}
\end{center}
\end{figure}

Figures \ref{heave_sml}  \& \ref{heave_big} show the predicted vertical force compared to measurements. The displacement has been subtracted from the results.   As the model ramps up to full speed, a mean suction force is induced on the model.   The oscillatory portion of the force is dominated by hydrostatics.  Once again, numerical predictions and laboratory measurements of the model running in the long wavelength waves are in good agreement.

%\begin{figure*}
%\begin{center}
%\begin{tabular}{lclc}
%(a)  \vspace{-7pt} & & (b)  \vspace{-7pt} & \\
%& \includegraphics[width=0.4\linewidth]{sml_heave.jpg} & & \includegraphics[width=0.4\linewidth]{big_heave.jpg} \\
%\end{tabular}
%\caption{\label{heave} Vertical (heave) force.  (a) $\lambda/L_o=1/2$. (b) $\lambda/L_o=2$.}
%\end{center}
%\end{figure*}

\subsection{Wave diffraction}

Another simulation was performed with a different grid density to predict the near-field wave disturbance.   A three-dimensional numerical simulation that uses 1536x256x256= 100,663,296 grid points, 24x4x4=384 sub-domains, and 384 nodes has been performed on a Cray XT3.  The length, width, depth, and height of the computational domain are respectively 4.0, 1.0, 1.0, 0.5 ship lengths ($L_o$).  Grid stretching is employed in all directions.  The smallest grid spacing is 0.0010  near the ship and mean waterline, and the largest grid spacing is 0.0060  in the far field.  The Froude number is $F_r=0.43$. One incident wavelength is considered:  $\lambda=2L_o$.   The wave steepness is $H/\lambda=0.03$, where $H=2 a$  is the wave height. The equations for the wavemaker are imposed ahead of the ship over the range $1\leq x \leq 1.5$. Initial transients are minimized by slowly ramping up the free-stream current and the incident wave amplitude.  The period of adjustment is $T_o=0.5$. For this simulation, the non-dimensional time step is $\Delta t$=0.00025.  The numerical simulation runs 28,000 time steps corresponding to 7 ship lengths.  The simulation requires about 90 hours of wall-clock time.

Figures \ref{nfaviews} (a \& b) show perspective views of the bow and stern waves.   A large plunging breaker forms near the bow.    At the stern, a large rooster tail forms.
Figures \ref{bowwaves} (a-h) show laboratory measurements compared to numerical predictions at evenly-spaced increments of the encounter wave period.   $\rm T_e$ denotes the encounter wave period.    The time increment between each comparison is $\rm T_e/8.$
The correlation coefficients between measurements and predictions  for t=0, $\rm T_e/8$, $\rm T_e/4$, $\rm 3T_e/8$, $\rm T_e/2$, $\rm 5T_e/8$, $\rm 3T_e/4$, and $\rm 7T_e/8$ are respectively 0.9193, 0.9414, 0.9612, 0.8737, 0.8191, 0.8900, 0.9347, and 0.8836.   
\section{Conclusions}

Using Model 5365, a 1/8.25 scale model of the R/V ATHENA,  a comprehensive data set of wave-induced drag and vertical forces, as well as the diffracted wave pattern  has been obtained at Froude numbers equal to 0.25 and 0.43.  This data set and the iges files of the trimmed model geometry are available to the international CFD community.  The wave-field perturbations due to the incoming wave field were measured using a Quantitative Free-Surface Visualization Laser Light Sheet technique, and compared to numerical predictions at evenly-spaced increments of the encounter wave period. Correlation coefficients between measurements and predictions varied from 82\% to 96\%.	

\section{Acknowledgements}

The Office of Naval Research supports this research with Dr. Patrick Purtell as the program manager.  At the Naval Surface Warfare Center this work is guided by Dr. Thomas Fu and Dr. Arthur Reed.  This work is supported in part by a grant of computer time from the DOD High Performance Computing Modernization Program (http://www.hpcmo.hpc.mil/).  The numerical simulations have been performed on the Cray XT3 at the U.S. Army Engineering Research and Development Center (ERDC).   The support of the flow visualization group at ERDC under the guidance of Paul Adams is gratefully acknowledged.

\begin{figure}
\begin{center}
\begin{tabular}{lc}
(a)  \vspace{-7pt} &\\
& \includegraphics[width=0.8\linewidth]{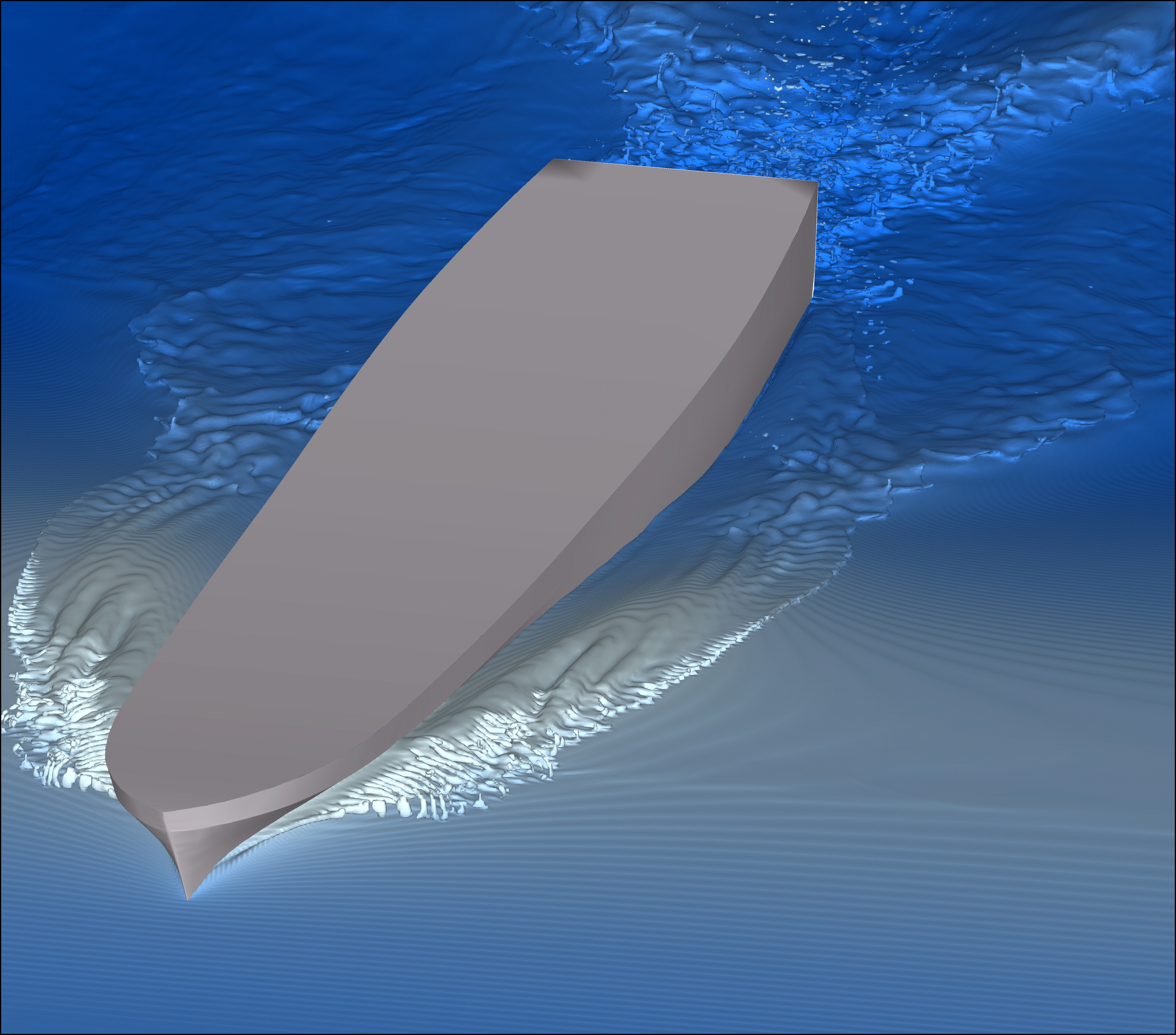} \\
(b)  \vspace{-7pt} & \\
 & \includegraphics[width=0.8\linewidth]{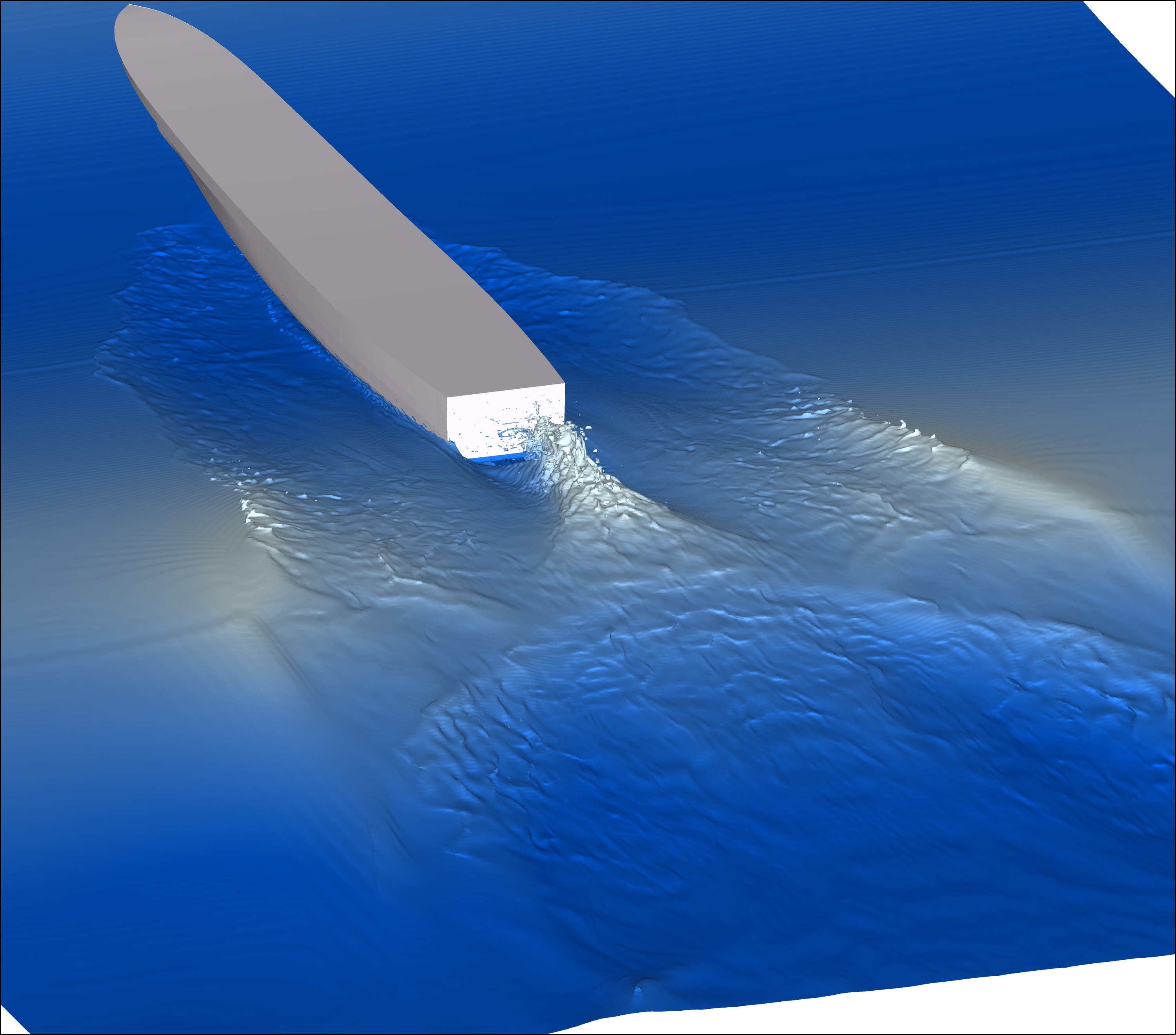} \\
\end{tabular}
\caption{\label{nfaviews} (a) Bow perspective view.    t=$\rm T_e/8.$.  (b) Stern perspective view.   t=$\rm 5T_e/8.$}
\end{center}
\end{figure}

\begin{figure*}
\begin{center}
\begin{tabular}[b]{lc}
(a) \vspace{-9pt} & \\
& \includegraphics[width=0.9\linewidth]{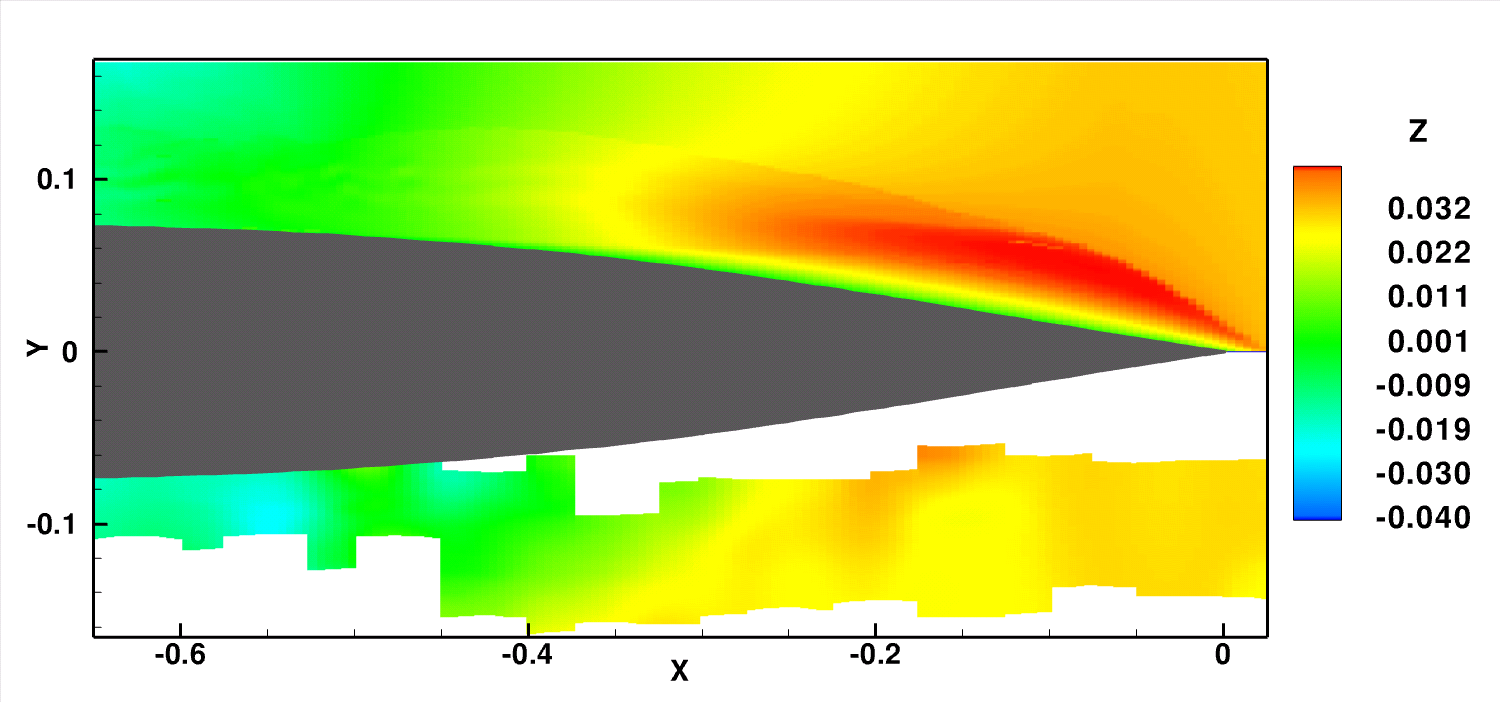} \\
(b) \vspace{-9pt} & \\
& \includegraphics[width=0.9\linewidth]{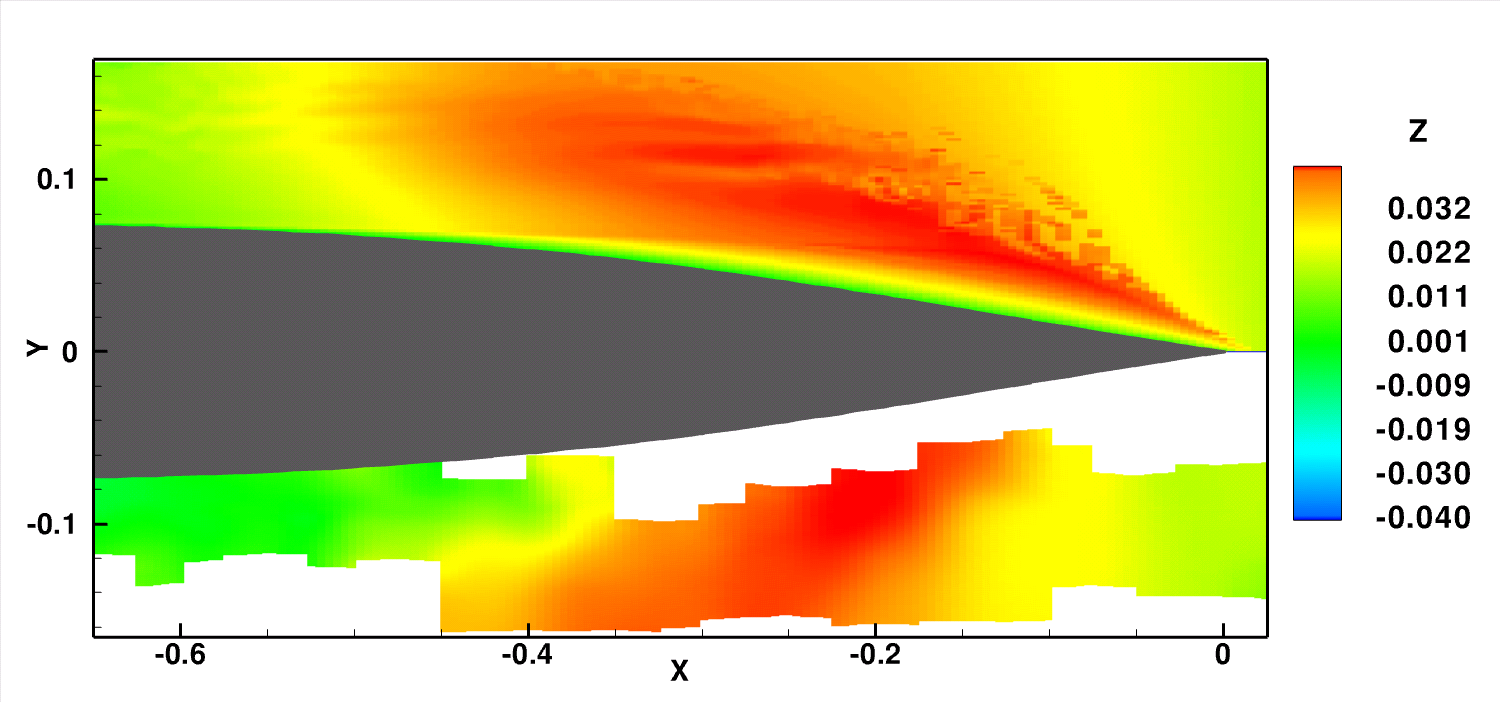} \\
(c) \vspace{-9pt} & \\
& \includegraphics[width=0.9\linewidth]{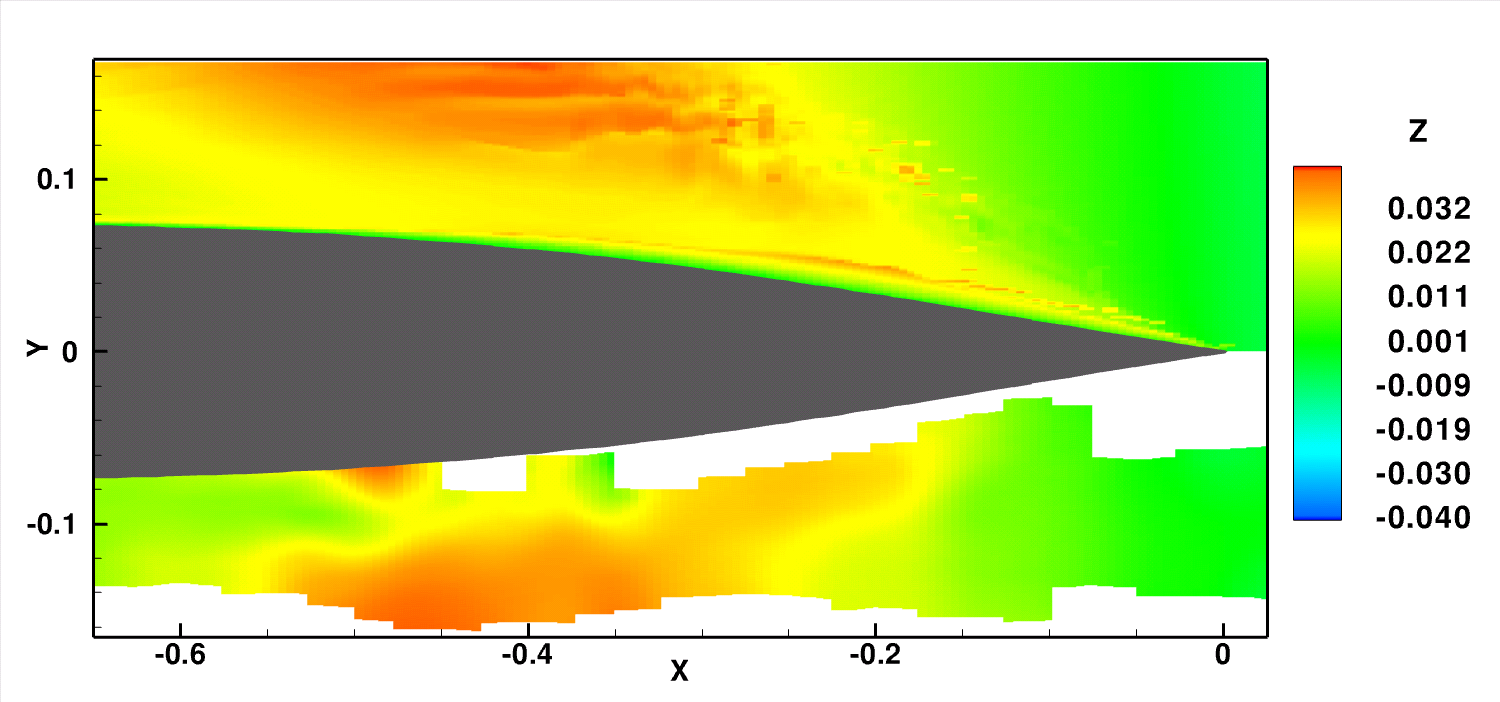} \\
\end{tabular}
\caption{\label{bowwaves} Diffracted wave near bow.  Experimental measurements are plotted at the bottom of each figure and numerical predictions are plotted at the top.   (a) t=0. (b) t=$\rm T_e/8.$ (c) t=$\rm T_e/4.$ }
\end{center}
\addtocounter{figure}{-1}
\end{figure*}

\begin{figure*}
\begin{center}
\begin{tabular}{lc}
(d) \vspace{-9pt} & \\
& \includegraphics[width=0.9\linewidth]{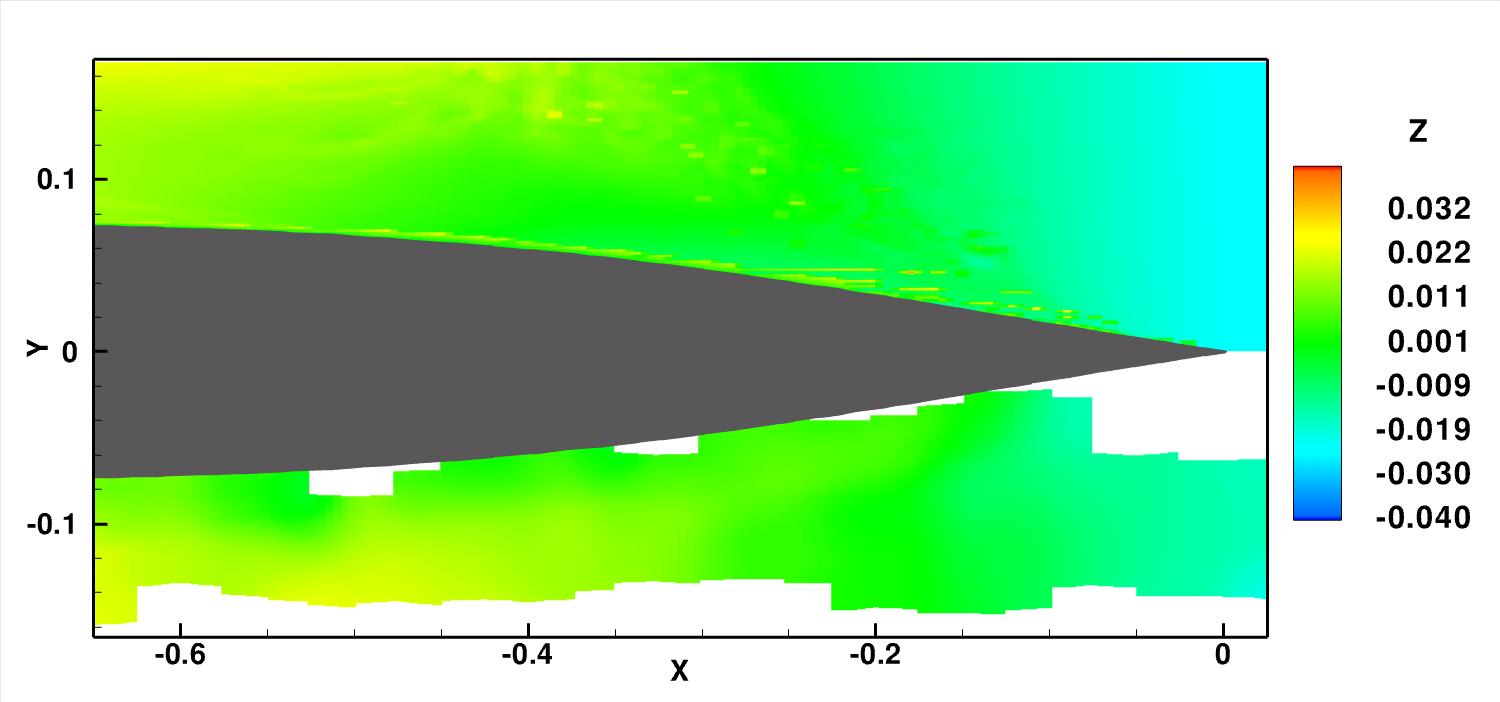} \\
(e) \vspace{-9pt} & \\
& \includegraphics[width=0.9\linewidth]{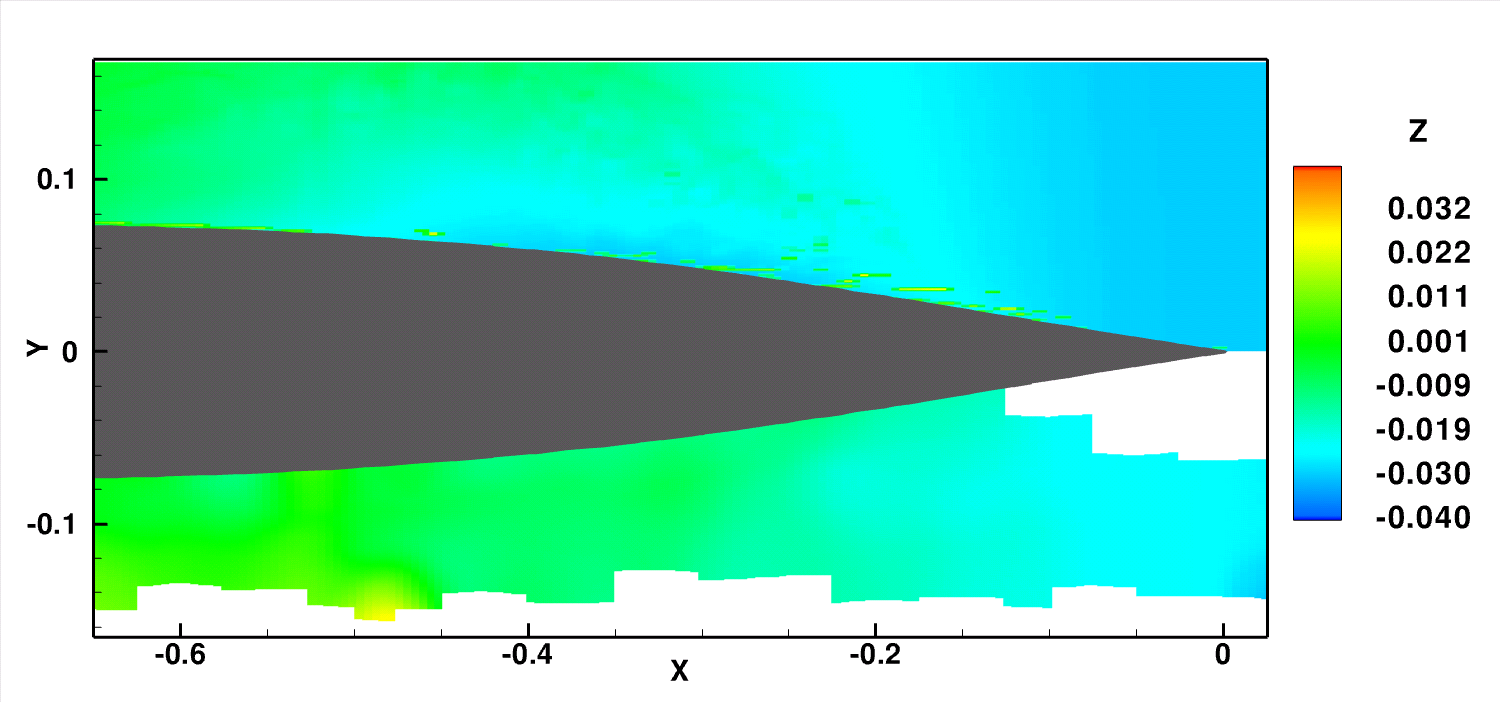} \\
(f) \vspace{-9pt} & \\
& \includegraphics[width=0.9\linewidth]{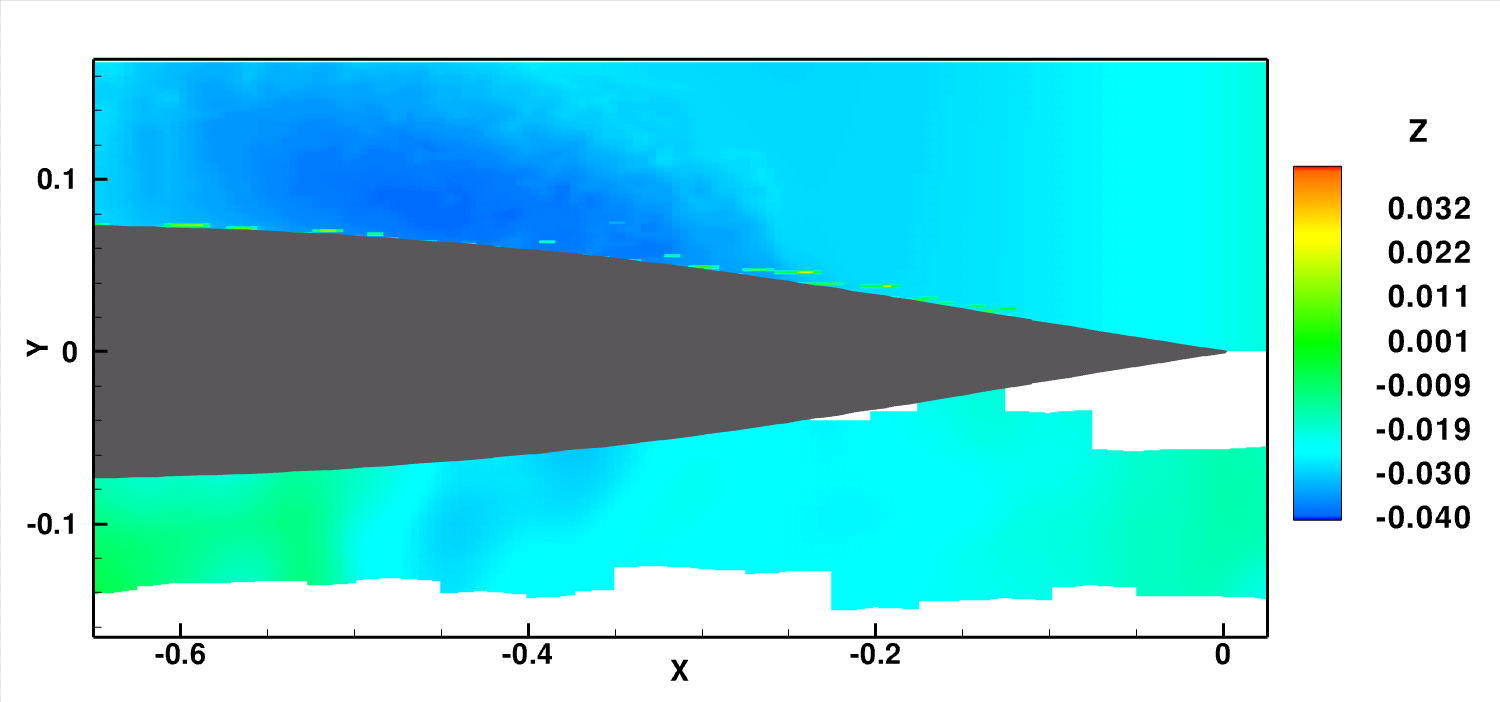}
\end{tabular}
\caption{Diffracted wave near bow, continued. (d) t=$\rm 3T_e/8.$ (e) t=$\rm T_e/2.$ (f) t=$\rm 5T_e/8.$}
\end{center}
\addtocounter{figure}{-1}
\end{figure*}

\begin{figure*}
\begin{center}
\begin{tabular}{lc}
(g) \vspace{-9pt} & \\
& \includegraphics[width=0.9\linewidth]{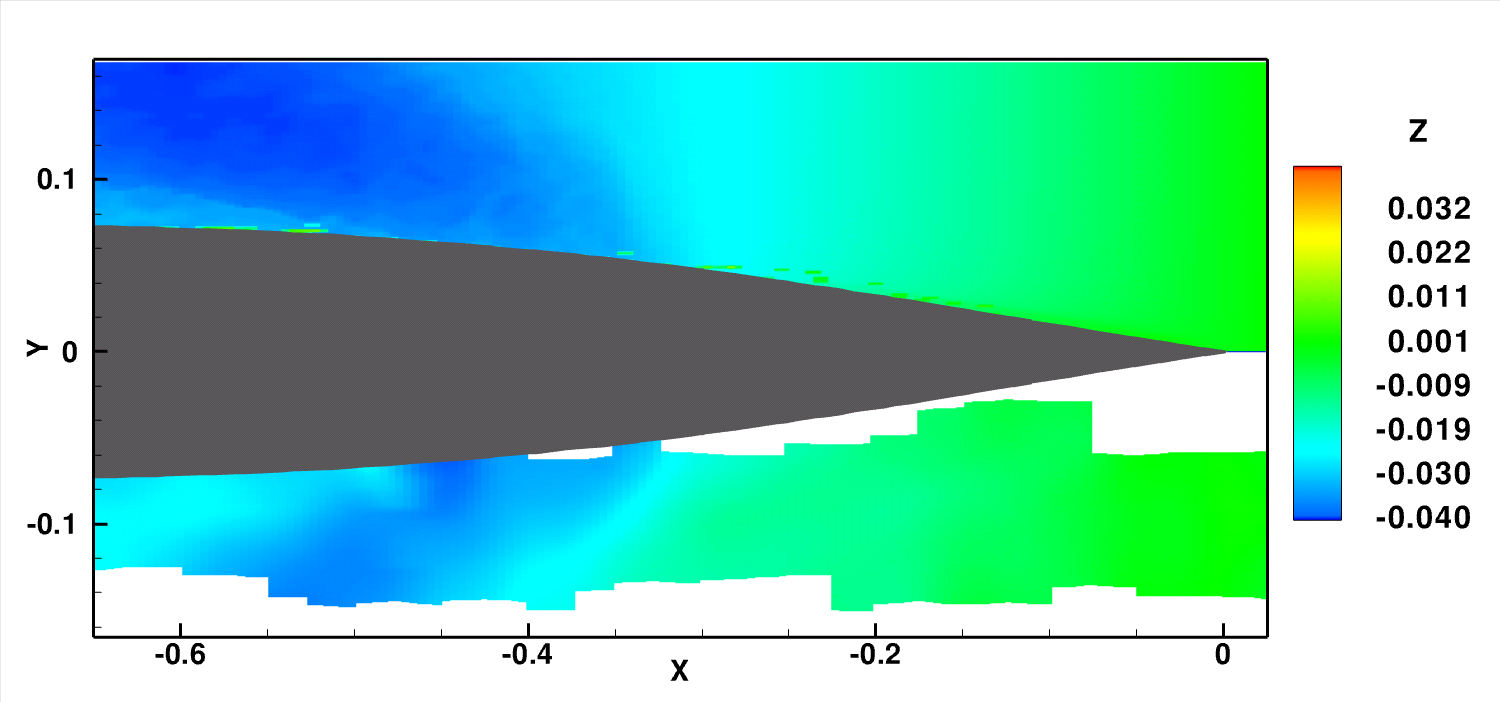} \\
(h) \vspace{-9pt} & \\
& \includegraphics[width=0.9\linewidth]{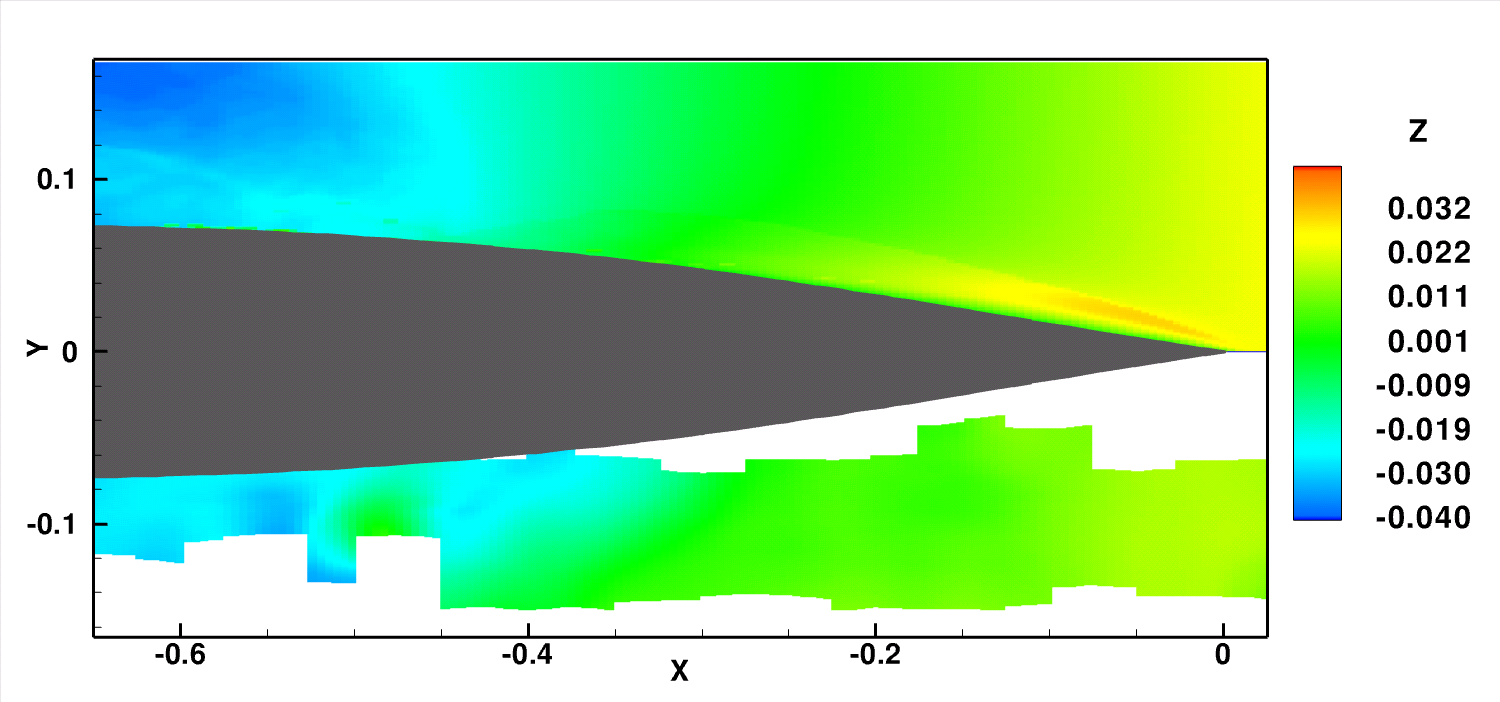}
\end{tabular}
\caption{Diffracted wave near bow, continued.  (g) t=$\rm 3T_e/4.$ (h) t=$\rm 7T_e/8.$}
\end{center}
\end{figure*}

\newpage

\bibliography{27onr}
\bibliographystyle{27onr}

\end{document}